\documentclass[prd,preprint,tightenlines,floatfix,showpacs,preprintnumbers,nofootinbib,eqsecnum]{revtex4}

 \usepackage[dvips,final]{graphicx}
  \usepackage{amssymb}
   \usepackage{amsmath}
    \usepackage{amsfonts}
     \usepackage{epsfig}
     \usepackage{color}
      \usepackage{bm} 

\usepackage{booktabs}
\usepackage{multirow}
\usepackage{slashed}
\usepackage{comment}
 \newcommand\ra{\rangle}
 \newcommand\beq{\begin{equation}}
 
 \newcommand\eeq{\end{equation}}
 \newcommand\beqn{\begin{eqnarray}}
 \newcommand\eeqn{\end{eqnarray}}

 \def\gsim{\mathrel{\rlap{\lower4pt\hbox{\hskip1pt$\sim$}}
 \raise1pt\hbox{$>$}}}
 \def\mb{\,\mbox{mb}}
 \def\fm{\,\mbox{fm}}
 
 \def\GeV{\,\mbox{GeV}}
 
 \def\lsim{\mathrel{\rlap{\lower4pt\hbox{\hskip1pt$\sim$}}
     \raise1pt\hbox{$<$}}}         
 \def\gsim{\mathrel{\rlap{\lower4pt\hbox{\hskip1pt$\sim$}}
     \raise1pt\hbox{$>$}}}         

 \def\beq{\begin{equation}}
 \def\eeq{\end{equation}}
 
 \def\beqy{\begin{eqnarray}} 
 \def\eeqy{\end{eqnarray}}
 \def\dfr{\mathrm{d}}
\def\Jpsi{J\!/\!\psi}
\def\psip{\psi'}
\def\Y{\Upsilon}
\def\Yp{\Upsilon'}
\def\Ypp{\Upsilon''}

\begin{document}


\title
{
Spin rotation effects in diffractive 
electroproduction\\ 
of heavy quarkonia
}

\author{Michal Krelina$^{1,2}$}
\email{michal.krelina@usm.cl}

\author{Jan Nemchik$^{2}$}
\email{jan.nemcik@fjfi.cvut.cz}

\author{Roman Pasechnik$^{3,4,5}$}
\email{roman.pasechnik@thep.lu.se}

\author{Jan \v Cepila$^{2}$}
\email{jan.cepila@fjfi.cvut.cz}

\affiliation{
{$^1$\sl
Departamento de F\'{\i}sica,
Universidad T\'ecnica Federico Santa Mar\'{\i}a,
Casilla 110-V, Valpara\'{\i}so, Chile
}\vspace{0.5cm}\\
{$^2$\sl 
Czech Technical University in Prague, FNSPE, B\v rehov\'a 7, 11519 
Prague, Czech Republic
}\vspace{0.5cm}\\
{$^3$\sl
Department of Astronomy and Theoretical Physics, Lund
University, SE-223 62 Lund, Sweden
}\vspace{0.5cm}\\
{$^4$\sl
Nuclear Physics Institute ASCR, 25068 \v{R}e\v{z}, Czech Republic
}\vspace{0.5cm}\\
{$^5$\sl Departamento de F\'isica, CFM, Universidade Federal 
de Santa Catarina, C.P. 476, CEP 88.040-900, Florian\'opolis, 
SC, Brazil\vspace{5mm}
}
}

\begin{abstract}
\vspace{5mm}In this work we present for the first time the comprehensive study of the Melosh spin rotation effects in diffractive electroproduction of $S$-wave heavy quarkonia off a nucleon target. Such a study has been performed within the color dipole approach using, as an example and a reference point, two popular parametrizations of the dipole cross section and two potentials describing the interaction between $Q$ and $\bar Q$ and entering in the Schr\"odinger equation based formalism for determination of the quarkonia wave functions. We find a strong onset of spin rotation effects in $1S$ charmonium photoproduction which is obviously neglected in present calculations of corresponding cross sections. For photoproduction of radially excited $\psip(2S)$ these effects are even stronger leading to an increase of the photoproduction cross section by a factor of $2\div 3$ depending on the photon energy. Even in production of radially excited $\Yp(2S)$ and $\Ypp(3S)$ they can not be neglected and cause the $20-30\,\%$ enhancement of the photoproduction cross section. Finally, we predict that the spin effects vanish gradually with photon virtuality $Q^2$ following universality properties in production of different heavy quarkonia as a function of $Q^2 + M_V^2$.
\end{abstract}


\maketitle
%
%
%
\section{Introduction}
\label{Sec:Intro}
%
%
%
Detailed investigation of elastic virtual photo- and electroproduction of heavy 
quarkonia $\gamma^*\,p\to\Jpsi(\psip,\Y,\Yp,...)\,p$ is presently a very relevant 
and dynamic research area. Indeed, the charmonium (bottomonium) suppression phenomenon 
in nucleus-nucleus interactions represents an alternative probe for the medium created 
in heavy-ion collisions (HICs) \cite{satz}. Since the size of heavy 
quarkonia is relatively small, the corresponding calculations of the production amplitudes are expected to be mainly based on perturbative QCD (pQCD), thus, minimizing the uncertainties coming from the nonperturbative interactions (for a detailed review on quarkonia physics, see e.g.~Refs.~\cite{ivanov-04,brambilla-04,brambilla-10}).

This expectation can be verified within the color dipole approach \cite{jan-94}.
Here, at large photon virtualities, $Q^2\gg M_V^2$, where $M_V$ is the mass of 
a vector meson, the small-size behavior of the dipole cross section \cite{zkl}, 
$\sigma_{\bar qq}(r)\propto r^2$, together with the shrinkage of the $Q\bar Q$ component 
of the photon with $Q^2$, lead to the effect known as the \textit{scanning phenomenon} 
\cite{bzk-93,bzk-94,jan-94,jan-96,jan-01,jan-07}. Namely, the vector meson production amplitude is scanned predominantly at dipole sizes of the order of $r \sim r_S$,
in terms of the \textit{scanning radius} $r_S$ given by
%
\beqn
r_S 
\approx
\frac{3}{\varepsilon} =  
\frac{Y}{2\sqrt{Q^{2}\,z\,(1-z) + m_Q^{2}}}
\sim
\frac{Y}{\sqrt{Q^2 + M_V^2}}\, ,
\label{scan-rad}
\eeqn
%
where $m_Q$ is the mass of heavy quark and $z$ is the fraction of the photon momentum carried by a quark or antiquark from the $Q\bar Q$ component of the photon representing
the dipole of transverse separation $\vec r$.

The scanning phenomenon allows to study the transition from the nonperturbative region 
of large $r_S$ to the perturbative region of very small $r_S$ by means of variation of $Q^2$ and the mass of vector meson $M_V$. The onset of pQCD regime requires that $r_S\lsim r_0$ \cite{jan-07}, where $r_0\sim 0.3\,\fm$ represents the gluon propagation radius \cite{spots,drops}. This leads to the following condition \cite{jan-07},
%
 \beqn
(Q^2 + M_V^2)\gsim Q_{\rm pQCD}^2 = \frac{Y^2}{r_0^2}\, .
\label{pqcd}
\eeqn
%
Thus the charmonium real photoproduction ($Q^2=0$) does not probe the perturbative region and one needs $Q^2\gsim 10\div 20\,\GeV^2$ to enter the pQCD regime. This is not so true 
for the bottomonium electroproduction where pQCD calculations can be safely 
performed already in the photoproduction limit $Q^2\to 0$ due to a large bottom mass.

On the other hand, another manifestation of the scanning phenomenon is that
the fixed values of $r_S$ at different $Q^2$ lead to very similar cross sections
in electroproduction of different vector mesons, i.e. the onset of such universality 
appears at the same values of $Q^2 + M_V^2$.

Due to the nodal structure of wave functions \cite{bzk-93,jan-94-2s,jan-96,jan-00a,jan-00b,jan-07}, for production of radially 
excited vector mesons the scanning phenomenon should be treated at sufficiently large 
scales $Q^2$
%
 \beqn
(Q^2 + M_V^2)\gg \frac{Y^2}{r_n^2}\, ,
\eeqn
%
where $r_n$ represents the position of the first node in the corresponding wave functions.

Within the nonrelativistic approximation ($M_V\sim 2 m_Q$, $z\sim 0.5$) the
factor $Y$ in Eq.~(\ref{scan-rad}) was evaluated in Ref.~\cite{jan-94} to be 
$Y\approx 6$
%
and corresponds to a contribution from the symmetric $Q\bar Q$ photon configurations to the quarkonium production, when the quark $Q$ and antiquark $\bar Q$ carry almost the same fraction of the meson momentum, $z\sim 0.5$.  
%
Such an approximation has a reasonable accuracy for charmonium 
and is rather good for bottomonium production. In general, the parameter $Y$ 
depends on polarization and increases slowly with $Q^2$ \cite{jan-96,jan-07} 
due to relativistic corrections.
%
Here the $Q^2$ dependence of $Y$ comes from the large-size asymmetric $Q\bar Q$ configurations of the virtual photon when the quark or antiquark in the photon carry a very small or a very large fraction of the meson momentum, $z\to 0, 1$.
%

The above statement is such that elastic electroproduction of heavy quarkonia 
at $Q^2\gg 0$ probes mainly the phenomena at hard scale, when one should rely on pQCD calculations, while semihard-scale physics is of a nonperturbative origin and can be probed only at small $Q^2$ (see Eq.~(\ref{pqcd})). Given this statement, 
in the current paper we present for the first time a detailed analysis of effects of \textit{Melosh spin rotation} \cite{melosh} within the color dipole formalism  \cite{bzk-91,bzk-93,bzk-94,jan-94,jan-96,yura-00}. In this analysis, we revise the corresponding results from Ref.~\cite{yura-00} where we have found a missing factor
in expressions for the production amplitudes leading to a modification of results for elastic photo- and electroproduction of charmonia. Besides, we extend our study of spin rotation effects also to bottomonium production including the excited $(2S)$ and $(3S)$  
states.

In the next Sec.~\ref{Sec:col-dip}, we start with a short review of the factorised light-cone dipole formalism to elastic photo- and electroproduction of heavy quarkonia.
The corresponding generic formula for the forward amplitude introduced in Sect.~\ref{Sec:dipole-formula} depends on our knowledge of several ingredients such as 
the light-cone distribution or the wave function of a transversely or longitudinally polarised virtual photon for a $Q\bar Q$ fluctuation, the light-cone wave function of a heavy quarkonia, and  the flavor independent universal dipole cross section depending on energy and the transverse separation between $Q$ and $\bar Q$.

The Sec.~\ref{Sec:photon} is devoted to description of the well-known perturbative distribution of heavy quarks in the photon. Here, we include the Melosh spin transformation and obtain the final expressions for photo- and electroproduction amplitudes at different polarizations modifying correctly the standard and frequently used formulas.

In Sec.~\ref{Sec:VM-wf} within the nonrelativistic approximation we describe how the corresponding wave functions for heavy quarkonia in the rest $Q\bar Q$ frame can be obtained by solving the Schroedinger equation with different types of interquark $Q\bar Q$ potentials. 

Then in the next Sec.~\ref{Sec:QQ-wf} we describe and discuss the procedure based only on widely-used prescription allowing to perform a Lorentz boost to the light-front frame. 

As the last ingredient of the production amplitude, in Sec.~\ref{dipole-cs} we introduce two parametrizations of the dipole cross section used in our calculations.

The results of our model parameter-free calculations are compared with available data for $\Jpsi(1S)$ and $\Y(1S)$ photoproduction in Sec.~\ref{Sec:results}, and a reasonable agreement has been found. We also demonstrate a strong onset of spin rotation effects, which are usually neglected in present calculations, in production of radially excited $2S$ and $3S$ states. They lead to an enhancement of the ratio of $\psip(2S)$ to $\Jpsi(1S)$ photoproduction cross sections by a factor of $2\div 3$ in a reasonable agreement with the data. In electroproduction of charmonia the effect of the Melosh spin transformation gradually decreases with photon virtuality manifesting a weak onset in bottomonium production as a clear consequence of the scanning phenomenon discussed above.

Finally, the results of the paper are summarized in Sec.~\ref{Sec:conclusions}. Here we discuss how manifestations and magnitudes of spin rotation effects are correlated with production of different states of heavy quarkonia.

%
%
%
\section{Color-dipole approach for electroproduction of heavy quarkonia}
\label{Sec:col-dip}
%
%
%

In the framework of color dipole approach \cite{bzk-91,bzk-93,bzk-94,jan-94,jan-96,yura-00},
the projectile photon (real or with virtuality $Q^2$) undergoes strong interactions via its Fock states containing quarks and gluons with the proton target in the rest frame 
of the target.

The Fock state expansion for the relativistic vector meson starts with the quark-antiquark $Q\bar Q$ state which can be considered as a color dipole. Here the relevant variables are the dipole moment $r$, which is the transverse separation between the quark and antiquark and $z$ representing the fraction of the light-cone (LC) momentum of the meson carried by a quark. The interaction of the relativistic color dipole with the target nucleon is described by the universal flavor independent color dipole cross section (introduced for the first time in Ref.~\cite{zkl}), which is a universal function of $r$ and energy.

However, at high energy one should include also contributions of higher Fock states containing gluons, such as $Q\bar Q+g$ etc. Here in the leading-$\log(1/x)$ approximation
the effect of higher Fock states can be reabsorbed into the energy dependence of the dipole cross section. Once the dipole scattering occurs, a coherent $Q\bar Q$ state forms a vector meson with given quantum numbers.

%
%
%
\subsection{Dipole formula for production amplitude}
\label{Sec:dipole-formula}
%
%
%

The forward amplitude for exclusive electroproduction of a vector meson $V$ with mass $M_V$ in the target rest frame can be treated in the quantum-mechanical formalism
\cite{bzk-91,bzk-93,bzk-94,jan-94,jan-96,yura-00},
%
\beqn
\mathrm{Im}\,\mathcal{A}^{\gamma^{\ast}p\rightarrow Vp}_{T,L}(x,Q^2)
=
\int\dfr^{2}r\,\int\limits_{0}^{1}\dfr z \,\Psi^{\dagger}_{V}(r,z)\,
\Psi_{\gamma^{\ast}_{T,L}}(r,z;Q^2)\,\sigma_{q\bar q}(x,r)
\label{prod-amp}
\eeqn
%
with the normalization
%
\beqn
\left.\frac{d\sigma^{\gamma^{\ast}p\rightarrow Vp}}{dt}\right|_{t=0}
= 
\frac{|\mathcal{A}|^2}{16\,\pi}\,.
\eeqn
%

In Eq.~(\ref{prod-amp}) $x$ is the standard Bjorken variable defined at small $Q^2$ by prescription from Ref.~\cite{ryskin-95} as $x=(M_V^2+Q^2)/s$ with the c.m. energy squared $s$ of the electron-proton system; $\Psi_{V}(r,z)$ is the LC vector meson wave function; 
$\Psi_{\gamma^{\ast}_{T,L}}(r,z;Q^2)$ is the LC distribution or the wave function of a transversely (T) or longitudinally (L) polarized virtual photon for a $Q\bar Q$ fluctuation; $\vec r$ is the transverse size of the $Q\bar Q$ dipole; and $z=p_Q^+/p_{\gamma}^+$ is the boost-invariant fraction of the photon momentum 
carried by a heavy quark (or antiquark). The universal dipole cross section $\sigma_{q\bar q}(x,r)$ describes a QCD interaction of the $Q\bar Q$ dipole 
(with transverse separation $r$) with the target.

As was already discussed in Sec.~\ref{Sec:Intro}, in the analysis of heavy quarkonia production in the considering regime one can safely use the nonrelativistic QCD 
(NRQCD) limit with a good accuracy and neglect the relative motion of $Q$ and 
$\bar Q$ such that the momentum fraction is $z\sim 1/2$ and the corresponding
LC wave function reduces down to $\Psi_{V}(r,z)\propto \delta(z-1/2)$.

%
%
%
\subsection{The $Q\bar Q$ wave function of the photon}
\label{Sec:photon}
%
%
%

The perturbative distribution amplitude (or ``wave function'', in what follows) for
the $Q\bar Q$ Fock component of the photon is well known \cite{kogut,bjorken,nnn}, 
and for $T$ and $L$ polarized photons it has the following form,
%
\beqn
\Psi^{(\mu,\bar\mu)}_{\gamma^{\ast}_{T,L}}(r,z;Q^2)
=
\frac{\sqrt{N_c\alpha_{\rm em}}}{2\pi} Z_Q\,\chi_Q^{\mu\dag}
\hat{\mathcal{O}}_{T,L}\tilde\chi_{\bar Q}^{\mu}\,K_0(\varepsilon r) \,,
\label{gam-wf}
\eeqn
%
where $N_C = 3$ is the number of colors, $\varepsilon^2=z(1-z)Q^2+m_Q^2$, $Z_Q$ is
the electric charge of the heavy quark ($Z_c=2/3$, $Z_b = 1/3$), $K_0(\varepsilon r)$ is the modified Bessel function of the second kind, and $\chi^\mu_Q$ and $\tilde\chi^{\bar\mu}_{\bar Q}\equiv i\sigma_y\chi^{\bar\mu\ast}_{\bar Q}$ are the two-component spinors of the quark and antiquark in the light-front frame 
normalized as \cite{bzk-dy-01}
%
\beqn
\sum\limits_{\mu,\bar \mu}\left(\chi^{\mu\dag}_Q\hat A
\tilde\chi^{\bar\mu}_{\bar Q}\right)^\ast\left(\chi^{\mu\dag}_Q\hat B
\tilde\chi^{\bar\mu}_{\bar Q}\right)
=
\mathrm{Tr}(\hat A^\dag\hat B)\, .
\eeqn
%

The operators $\hat{\mathcal{O}}_{T,L}$ in Eq.~(\ref{gam-wf}) read,
%
\beqn
\mathcal{O}_T 
&=& 
m_Q\vec\sigma\cdot\vec e_\gamma + 
i(1-2z)(\vec\sigma\cdot\vec n)(\vec e_\gamma\cdot\vec\nabla_r) +
(\vec n\times\vec e_\gamma)\vec\nabla_r \,, \nonumber\\
\mathcal{O}_L 
&=& 
2 Q z(1-z)\vec\sigma\cdot \vec n \,, \qquad \vec\sigma
=
(\sigma_x,\sigma_y,\sigma_z) \,, \qquad \vec\nabla_r 
\equiv \partial/\partial \vec{r} \,,
\eeqn
%
where $\vec e_\gamma$ is the transverse photon polarisation vector, $\vec n=\vec p_{\gamma}/|\vec p_{\gamma}\,|$ is a unit vector along the photon momentum, and $\sigma_{x,y,z}$ are the Pauli matrices. 

In order to incorporate the spin rotation effects in photo- and electroproduction of heavy quarkonia, we assume a simple factorization of the spatial and spin-dependent parts of the vector meson wave function such as
%
\beqn
\Psi_{V}^{(\mu,\bar\mu)}(z,\vec p_T)
=
U^{(\mu,\bar\mu)}(z,\vec p_T)\Psi_{V}(z,p_T) \,,
\label{PsiQQ}
\eeqn
%
where the operator
%
\beqn
U^{(\mu,\bar\mu)}(z,\vec p_T)
=
\frac{1}{\sqrt{2}}\xi_Q^{\mu\dag}\vec\sigma
\vec e_{V}\tilde\xi_{\bar Q}^{\bar \mu}\,,\qquad 
\tilde\xi_{\bar Q}^{\bar\mu}
=
i\sigma_y\xi_{\bar Q}^{\bar\mu\ast} \,, 
\label{Uini}
\eeqn
%
is expressed in terms of the vector meson polarisation vector $\vec e_{V}$ and quark spinors $\xi$ in the meson rest frame. The latter are related to spinors $\chi$ in the light-front description by the following relation,
%
\beqn
\xi^\mu_Q
=
R(z,\vec p_T)\chi_Q^\mu \,, 
\qquad \xi_{\bar Q}^{\bar\mu}
=
R(1-z,-\vec p_T)\chi_{\bar Q}^{\bar\mu} \,,
\label{spinrot}
\eeqn
%
which is known as the Melosh spin rotation \cite{melosh,yura-00} with the corresponding
$R$-matrix given by
%
\beqn
R(z,\vec p_T)
=
\frac{m_Q + z M_V - i(\vec\sigma\times\vec n)\vec p_T}
{\sqrt{(m_Q+z\,M_V)^2+p_T^2}} \,.
\label{melosh}
\eeqn
%

Using the quarkonium wave function given by Eq.~(\ref{PsiQQ}) we assume that the vertex $\Psi \to Q\bar{Q}$ differs from the photon-like $\gamma^{\ast} \to Q\bar{Q}$ vertex 
with the structure $\psi_\mu \bar{u} \gamma^\mu u $ used in Refs.~\cite{ryskin-92,brodsky-94,frankfurt-95,jan-96}. As was noticed in Ref.~\cite{yura-00} for the case of charmonium production, assuming a similar structure of both vertices, the corresponding wave function in the $c\bar c$ rest frame 
contains both $S$- and $D$-wave states. However, the weight of the $D$-wave
is strongly correlated with the structure of the vertex and cannot be proved by any reasonable nonrelativistic $c\bar c$ interaction potential.

Substituting
%
\beqn
\tilde\xi_{\bar Q}^{\bar\mu}
=
i\sigma_yR^\ast(1-z,-\vec p_T)(-i)\sigma_y^{-1}
\tilde\chi_{\bar Q}^{\bar\mu} \,, \qquad
\xi_Q^{\mu\dag}=\chi_Q^{\mu\dag}R^{\dag}(z,\vec p_T) \,,
\eeqn
%
into Eq.~(\ref{Uini}) one gets finally
%
\beqn
U^{(\mu,\bar\mu)}(z,\vec p_T)
=
\frac{1}{\sqrt{2}}\chi_Q^{\mu\dag}R^\dag(z,\vec p_T)
\vec\sigma\cdot\vec e_{V}\sigma_y R^\ast (1-z,-\vec p_T)
\sigma_y^{-1}\tilde\chi_{\bar Q}^{\bar \mu} \,.
\eeqn
%
Then the resulting dipole formula for the amplitude of photo- and electroproduction of heavy quarkonia reads,
%
\beqn
\mathrm{Im}\mathcal{A}^{\gamma^{\ast}p\rightarrow Vp}_{T,L}(x,Q^2)
=
\int\limits_0^1\dfr z \int\dfr^2r \Sigma_{T,L}(z,\vec r;Q^2)\,
\sigma_{q\bar q}(x,r)\, ,
\label{sr1}
\eeqn
%
where
%
\beqn
\Sigma_{T,L}(z,\vec r;Q^2)
=
\int\frac{\dfr^2 p_T}{2\pi}e^{-i\vec p_T\vec r}
\Psi_{V}(z,p_T)\sum\limits_{\mu,\bar\mu}U^{\dagger(\mu,\bar\mu)}(z,\vec p_T)
\Psi^{(\mu,\bar\mu)}_{\gamma^{\ast}_{T,L}}(r,z;Q^2)\,.
\label{sr2}
\eeqn
%

Finally, using Eqs.~(\ref{sr1}) and (\ref{sr2}) one can arrive at the following final expressions for photo- and electroproduction amplitudes of heavy quarkonia in 
the polarised photon-nucleon scattering (compare with formulas in Ref.~\cite{yura-00}),
%
\beqn
&& \mathrm{Im}\mathcal{A}_{L}(x,Q^2)
=
\int\limits_0^1\dfr z\int\dfr^2\,r\,\Sigma_{L}(z,r;Q^2)\,
\sigma_{q\bar q}(x,r) \,,  
\label{AL}
\\
&& \Sigma_{L}(z,r;Q^2)
= 
Z_q\,\frac{\sqrt{N_c\alpha_{em}}}{2\pi\sqrt{2}}
\,4 Q z (1-z) K_0 (\varepsilon r)\int p_T\dfr p_T\,J_0(p_T\,r)
\Psi_{V}(z,p_T)\frac{m_T m_L + m_Q^2}{m_Q(m_T+m_L)} 
\,, \nonumber
\eeqn
%
for a longitudinally polarised photon\footnote{Here, we have found an additional factor of $\sqrt{2}$ which was missed in calculations in Ref.~\cite{yura-00}.}, and
%
\beqn
&& \mathrm{Im}\mathcal{A}_{T}(x,Q^2)
=
\int\limits_0^1\dfr z\int\dfr^2r 
\left[\Sigma^{(1)}_{T}(z, r;Q^2)\sigma_{q\bar q}(x,r) 
+
\Sigma^{(2)}_{T}(z, r;Q^2)
\frac{\dfr \sigma_{q\bar q}(x,r)}{\dfr r}\right]\,,   
\label{AT}
\\
&& \Sigma^{(1)}_{T} 
= 
Z_q\,\frac{\sqrt{N_c\alpha_{em}}}{2\pi\sqrt{2}}\,
2 K_0(\varepsilon r)\int \dfr p_TJ_0(p_T r)\Psi_{V}(z,p_T)
p_T\,\frac{m_T^2+m_Tm_L-2p_T^2z(1-z)}{m_T+m_L} 
\,, \nonumber \\
&& \Sigma^{(2)}_{T} 
= 
Z_q\,\frac{\sqrt{N_c\alpha_{em}}}{2\pi\sqrt{2}}\,
2 K_0(\varepsilon r)\int \dfr p_T J_1(p_T r)
\Psi_{V}(z,p_T)\frac{p_T^2}{2}\,
\frac{m_T+m_L+m_T(1-2z)^2}{m_T(m_T+m_L)} 
\,, \nonumber
\eeqn
%
for a transversely polarised photon. In above formulas,
%
\beqn
m_T^2 = m_Q^2 + p_T^2 \,, 
\qquad m_L^2 = 4m_Q^2\,z(1-z) \,,
\eeqn
%
such that the meson mass squared is
%
\beqn
M_V^2 = \frac{m_T^2}{z(1-z)} \,.
\eeqn
%

Consequently, the total integrated electroproduction $\gamma^{\ast}p\rightarrow Vp$ cross section is conventionally represented as a sum of $T$ and $L$ contributions \cite{yura-00}
%
\beqn
\sigma^{\gamma^{\ast}p\rightarrow Vp}(x,Q^2)
=
\sigma^{\gamma^{\ast}p\rightarrow Vp}_{T} +
\tilde{\varepsilon}\,
\sigma^{\gamma^{\ast}p\rightarrow Vp}_{L}
=
\frac{1}{16\pi B}\left(\Big\vert 
\mathcal{A}^{\gamma^{\ast}p\rightarrow Vp}_{T}\Big\vert^{2}+
\tilde{\varepsilon}\Big\vert \mathcal{A}^{\gamma^{\ast}p\rightarrow Vp}_{L}
\Big\vert^{2}\right) \,,
\label{final}
\eeqn
%
where $B$ is the slope parameter fitted to data and $\tilde{\varepsilon}$ represents the photon polarization with the value $\tilde{\varepsilon} = 0.99$ obtained from H1 HERA data \cite{h1-00}.

Since the relative contribution of effects of the Melosh spin rotation to the integrated cross section of elastic electroproduction of heavy quarkonia does not depend
on the magnitude of the slope parameter, so for simplicity, in all calculations,
we take a constant experimental value $B=4.73$ GeV$^{-2}$ \cite{h1-00} as in Ref.~\cite{yura-00}.

Note that the amplitudes $\mathcal{A}_{T}$ and $\mathcal{A}_{L}$ in Eq.~(\ref{final}) include also the corrections for the real part \cite{bronzan-74,jan-96,forshaw-03},
%
\beqn
{\mathcal A}^{\gamma^{\ast}p\rightarrow Vp}_{T,L}(x,Q^2) 
= 
{\mathrm{Im}\mathcal{A}}^{\gamma^{\ast}p\rightarrow Vp}_{T,L}(x,Q^2)
\,
\left(1 - i\,\frac{\pi}{2}\,\frac
{\partial
 \,\ln\,{\mathrm{Im}\mathcal{A}}^{\gamma^{\ast}p\rightarrow Vp}_{T,L}(x,Q^2)}
{\partial\,\ln x} \right)\ .
  \label{re/im}
\eeqn
%

%
%
%
\subsection{Quarkonium wave function}
\label{Sec:VM-wf}
%
%
%

The quarkonia wave functions in the $Q\bar Q$ rest frame and in impact parameter representation have been obtained by solving the Schroedinger equation for different 
interaction potentials between $Q$ and $\bar Q$ following the procedure from \cite{yura-00}. The corresponding Schroedinger equation has the following form,
%
\beqn
\left(-\frac{\Delta_r}{m_Q}+V_{Q\bar Q}(r)\right)\Psi_{nlm}(\vec r)
=
E_{nl}\Psi_{nlm}(\vec r)\,
\eeqn
%
where the wave function $\Psi_{nlm}(\vec r)$ depends on three-dimensional $Q\bar Q$ separation $\vec r$ and can be expressed in the factorized form,
%
\beqn
\Psi_{nlm}(\vec r)=\Psi_{nl}(r)\cdot Y_{lm}(\theta,\varphi)
\label{nrel-wf}
\eeqn
%
with $\Psi_{nl}(r)$ and $Y_{lm}(\theta,\varphi)$ representing the radial and orbital parts of the wave function. Consequently, the Schroedinger equation for the radial
part $\Psi_{nl}(r)$ was solved numerically.

In the present paper in our comprehensive study of the onset of Melosh spin rotation effects in photo- and electroproduction of heavy quarkonia we adopt only two heavy quark interaction potentials $V_{Q\bar Q}(r)$, which provide the best description of available data. Besides, the relative contribution of spin rotation effects to integrated
cross sections of elastic processes, $\gamma^*\,p\rightarrow \Jpsi (\psip, \Y, \Yp, \Ypp)\,p$ is not correlated with our choice of a given potential. For this reason the following two interaction potentials $V_{Q\bar Q}(r)$ have been used:\\
\\
1. The Buchm\"uller-Tye potential (BT) \cite{bt-80}, which has a linear
string-like behavior at large transverse separations and
a Coulomb shape at small $Q\bar Q$-distances,
%
\beqn
V_{Q\bar Q}(r)
=
\frac{k}{r}-\frac{8\pi}{27}\frac{v(\lambda r)}{r} \,,
\eeqn
%
for $r\geq 0.01\,\fm$, and
%
\beqn
V_{Q\bar Q}(r)
=
- \frac{16\pi}{25}\frac{1}{r\ln\left(\frac{1}{\lambda^2_{\rm 
MS}r^2}\right)}\left(1+2\left(\gamma_E+\frac{53}{75}\right)
\frac{1}{\ln\left(\frac{1}{\lambda^2_{\rm MS}r^2}\right)}-\frac{462}{625}\frac{\ln\left( 
\ln\left(\frac{1}{\lambda^2_{\rm MS}r^2}\right)\right)}
{\ln\left(\frac{1}{\lambda^2_{\rm MS}r^2}\right)}\right) \,
\eeqn
%
for $r < 0.01\,\fm$. Here,
%
\beqn
\lambda_{\rm MS}=0.509\,\GeV\,,
\qquad k=0.153\,\GeV^2\,,
\qquad\lambda=0.406\,\GeV\,,
\eeqn
%
$\gamma_E=0.5772$ is the Euler constant, and the function $v(x)$ is provided 
numerically in Ref.~\cite{bt-80}. The corresponding masses of heavy quarks are 
the following: $m_c=1.48$GeV and $m_b=4.87$ GeV.\\
\\
2. The power-like (POW) potential \cite{barik-80}
%
\beqn
V_{Q\bar Q}(r)
=
- 6.41\,\GeV + (6.08\,\GeV)(r\cdot 1\,\GeV)^{0.106} \,,
\end{eqnarray}
with $m_c=1.334\,\GeV$ and $m_b=4.721\,\GeV$. 

%
%
%
\subsection{Light-cone quarkonium wave function}
\label{Sec:QQ-wf}
%
%
%

In order to calculate the amplitude (\ref{prod-amp}) of the elastic process
$\gamma^*\,p\to V\,p$ one needs to know the LC quarkonium wave function $\Psi_{V}(r,z)$. 
Similarly to the LC photon-quark wave function $\Psi^{T,L}_{\gamma^{\ast}}$, 
it is defined in the light-front frame. Here, the lowest Fock component $|Q\bar Q\ra$ is not related to the quarkonium wave function in the $Q\bar Q$ rest frame by applying a simple Lorentz boost. The solution of this problem is difficult, and without any unambiguous result up to now. One can find only recipes in the literature, and
we use one of them presented in Ref.~\cite{terentev}. Here, as the first step
we switch from the coordinate space to the momentum space applying the Fourier transformation to the nonrelativistic wave function (\ref{nrel-wf}),
$\Psi(\vec r)\Rightarrow \Psi(\vec p)$, where $p$ represents the quark 3-momentum.
Consequently, the wave function $\Psi(\vec p)$ should be appropriately boosted to the light-front frame. For this purpose, it can be expressed in terms of the $Q\bar Q$
invariant mass
%
\beqn
M_{Q\bar Q}^2 = 4(p^2+m_Q^2)\,,
\eeqn
%
with $p^2 = p_T^2 + p_L^2$, where $p_L$ is the longitudinal component of the quark 3-momentum $\vec p$. The same quantity in the light-front kinematics is given by
%
\beqn
M_{Q\bar Q}^2(p_T,z) = \frac{p_T^2+m_Q^2}{z(1-z)} \,.
\eeqn
%
The last two relations lead to following qualities,
%
\beqn
p^2=\frac{p_T^2+(1-2z)^2m_Q^2}{4z(1-z)} \,,
\label{boost1}
\eeqn
%
and
%
\beqn
p_L^2
=
\frac{(p_T^2+m_Q^2)(1-2z)^2}{4z(1-z)}
=
\biggl (z - \frac{1}{2}\biggr )^2 M_{Q\bar Q}^2(p_T,z)\, ,
\label{boost2}
\eeqn
%
which relate the kinematical variables in the infinite momentum and $Q\bar Q$ rest frames. As a result, this procedure leads to the following relation between the LC wave function $\Psi(p_T,z)$ and its counterpart in the $Q\bar Q$ rest frame $\Psi(p)$,
%
\beqn
\Psi(p_T,z)
=
\sqrt{2}\,
\frac
{(p^2+m_Q^2)^{3/4}}
{(p_T^2+m_Q^2)^{1/2}}\,
\Psi(p)\, ,
\eeqn
%
where $p=p(p_T,z)$ is given by Eq.~(\ref{boost1}). 

Note that in Ref.~\cite{bzk-boosting} the Terent'ev prescription \cite{terentev} 
for the Lorentz boosting presented above has been discussed and compared with 
the exact calculations obtained by using the sophisticated Green function approach.
Here, the boost-invariant Schroedinger equation for the Green function was derived,
which replaces its standard 2-dimensional LC form \cite{kst-99,jan-01}.
It was shown that for symmetric $c\bar c$ configurations $z\sim 0.5$
the corresponding LC wave functions for the ground-state charmonia practically
coincide with the results provided by the phenomenological prescription \cite{terentev}.

Finally, the LC wave function in the impact parameter representation is then 
obtained from $\Psi(p_T,z)$ performing the Fourier transformation,
%
\beqn
\Psi_{V}(r,z)
=
\int\limits_0^{\infty}\dfr p_T\, p_T J_0(p_T\,r)\,\Psi(p_T,z)\,.
\eeqn
%

%
%
%
\subsection{Dipole cross section}
\label{dipole-cs}
%
%
%

The essential ingredient of the dipole formalism introduced in Ref.~\cite{zkl} 
is the dipole cross section $\sigma_{q\bar q}(r)$ with magnitudes at different
transverse separations $\vec r$ representing the eigenvalues of the elastic amplitude operator. So dipoles with a definite transverse size $\vec r$ are the eigenstates of the interaction in QCD originated as a result of the hadronic cross section expansion over them. This cross section has two main properties -- the flavor invariance due to universality of the QCD coupling, and the small size behavior, $\sigma_{q\bar q}(r) \propto r^2$ for $r\!\!\to\!0$, the property known as the \textit{color transparency}.
Thus the dipole cross section is a flavor independent universal function of $r$
and energy, and allows to describe in an uniform way various high-energy processes.

The hadron production cross sections are known to rise with energy, where the energy dependence can be included in two different ways corresponding to the same set of Feynman graphs. Within two-gluon exchange approximation \cite{zkl}, the dependence on energy comes from the higher-order corrections related to gluon radiation and the dipole cross section is constant. Another way is to involve higher Fock components containing gluons in addition to the $Q\bar Q$ state. Here we prefer to introduce the energy dependence 
via $\sigma_{q\bar q}(r,s)$ not including higher Fock states into the wave functions
(see Eq.~(\ref{prod-amp})).

In the limit of small dipole separations one can apply perturbative QCD results,
and the energy dependence comes as an effect of gluon radiation treated in the leading-$\log(1/x)$ approximation \cite{bfkl,book}. At large separations typical for light hadrons the effects of gluon bremsstrahlung can be also calculated taking the smallness of the quark-gluon correlation radius \cite{k3p}. However, we are interested also in the intermediate region of dipole sizes, which is more complicated. Here the dipole cross section still cannot be predicted reliably because of poorly known higher order pQCD corrections and nonperturbative effects. Therefore, we use a phenomenological form interpolating between the two limiting cases of small and large transverse separations. 

Although about ten different parametrizations of $\sigma_{q\bar q}(r,s)$ can be found recently in the literature, for our study we use only two of them since the relative contribution of spin rotation effects to elastic photo- and electroproduction cross sections of heavy quarkonia is not sensitive to our choice of $\sigma_{q\bar q}(r)$.

The first parametrization conventionally denoted as GBW has been suggested in Ref.~\cite{gbw} and has the following form,
\beqn
  \sigma_{q\bar q}(r,x)&=&23.03\left[1-e^{- r^2/r_0^2(x)}\right]\mb\ ,\\
  r_0(x) &=& 0.4 \left(\frac{x}{x_0}\right)^{\!\!0.144} \fm\ ,
  \nonumber
  \label{gbw}
\eeqn
where $x_0=3.04\cdot10^{-4}$. It well describes the data for DIS at small $x$, and at medium and high $Q^2$ \cite{gbw}. However, at small $Q^2$ it fails to describe the energy dependence of the hadronic total cross sections. Here, the Bjorken $x$ is not a well defined variable, and we prefer to use the following prescription known from Ref.~\cite{ryskin-95}, $x = (M^2_V + Q^2)/s = (M_V^2 + Q^2)/(W^2 + Q^2)$. 

A thorough analysis of this problem including a discussion of issues with the proper definition of $x$ has been presented in Ref.~\cite{kst-99}. Here, the dipole cross section denoted as KST contains an explicit dependence on the c.m. energy $\sqrt{s}$, rather than $x$, since $\sqrt{s}$ is a more appropriate variable for hadronic processes.
It has the following form \cite{kst-99}, which is similar to the one used 
in Ref.~\cite{gbw},
\beqn
  \sigma_{\bar qq}(r,s) &=& \sigma_0(s) \left[1 - e^{- r^2/r_0^2(s)}
  \right]\ .~~~~~~
  \label{kst}
\eeqn
Such a dipole cross section correctly reproduces the energy dependence of hadronic cross sections for the following choice
\beqn
  \sigma_0(s) &=& 23.6 \left(\frac{s}{s_0}\right)^{\!\!0.08}
  \left(1+\frac38 \frac{r_0^2(s)}{\left<r^2_{ch}\right>}\right)\mb\ ,\\
  r_0(s)      &=& 0.88 \left(\frac{s}{s_0}\right)^{\!\!-0.14}  \fm\ \,,
\eeqn
where $s_0 = 1000\GeV^2$, and the mean square of
the pion charge radius is $\left<r^2_{ch}\right>=0.44\fm^2$. 

An improvement of data description at large dipole separations leads to a somewhat 
worse description of data on the proton structure function at large $Q^2$. Apparently, the KST dipole cross section cannot provide the Bjorken scaling appropriately,
and the corresponding parameterization (\ref{kst}) can be applied successfully only up to $Q^2\approx 10\div 20\,\GeV^2$.

Since quarkonium production covers mainly the range of scales between the kinematic domains where either of these parametrizations is more successful than another, one can treat differences in predictions obtained by using the GBW (\ref{gbw}) and KST (\ref{kst}) dipole cross sections as a good measure of the underlined theoretical uncertainty. However, as was mentioned above, the choice of another arbitrary parametrization of the dipole cross section available in the literature has practically no impact on the relative magnitude of spin rotation effects, study of which represents the main scope of the present paper.

%
%
%
\section{Numerical results vs data}
\label{Sec:results}
%
%
%

As the first step, we investigate the universality in production of different heavy quarkonia as a function of $Q^2 + M_V^2$. As we have already discussed in Sec.~\ref{Sec:Intro}, this universality is controlled by the scanning radius $r_S$ (\ref{scan-rad}) where the $Q^2$ behavior of polarization-dependent scale factors $Y_{T,L}$ has been discussed in Refs.~\cite{jan-94,jan-96,jan-07} and is related to
the $Q\bar Q$ configurations when $Q$ or $\bar Q$ carries a dominant fraction of the photon momentum $z\to 1$. Such configurations are different for $T$ and $L$ polarized photons leading to a strong inequality $Y_{L} < Y_{T}$ as was analysed in Refs.~\cite{jan-94,jan-96,jan-07} assuming that the vertex $\Psi\rightarrow Q\bar Q$ has
the structure $\Psi_{\mu}\bar{u}\gamma_{\mu}u$ like for the leading photon Fock state $\gamma^*\rightarrow Q\bar Q$.

\begin{figure}[h]
\begin{center}
    \includegraphics[height=9.0cm]{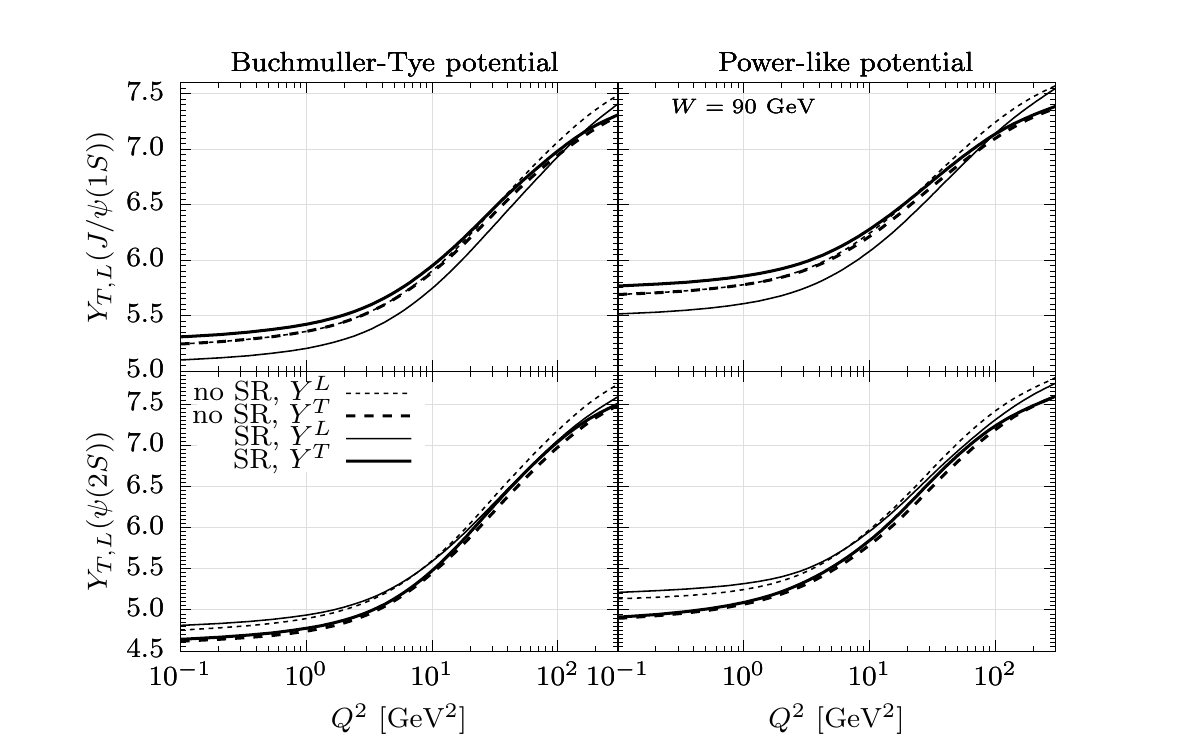}
    \caption{ \label{fig:psi-y}  
The $Q^2$ dependence of the scale parameters $Y_T$ and $Y_L$ defining the magnitude of the scanning radius, Eq.~(\ref{scan-rad}), is shown for production of $T$ and $L$ polarized $\Jpsi(1S)$ (upper boxes) and $\psip(2S)$ (lower boxes) vector mesons. The solid and dashed curves represent the values of $Y_T$ and $Y_L$ with and without effects
of Melosh spin rotation, respectively. The thin and thick curves correspond to production of $L$ and $T$ polarized vector mesons, respectively. The calculations have been performed with the vector meson wave function using the realistic Buchm\"uller-Tye \cite{bt-80} (left boxes) and power-like \cite{barik-80} (right boxes) potential.
}
\end{center}
\end{figure}

\begin{figure}[h]
\begin{center}
    \includegraphics[height=13.0cm]{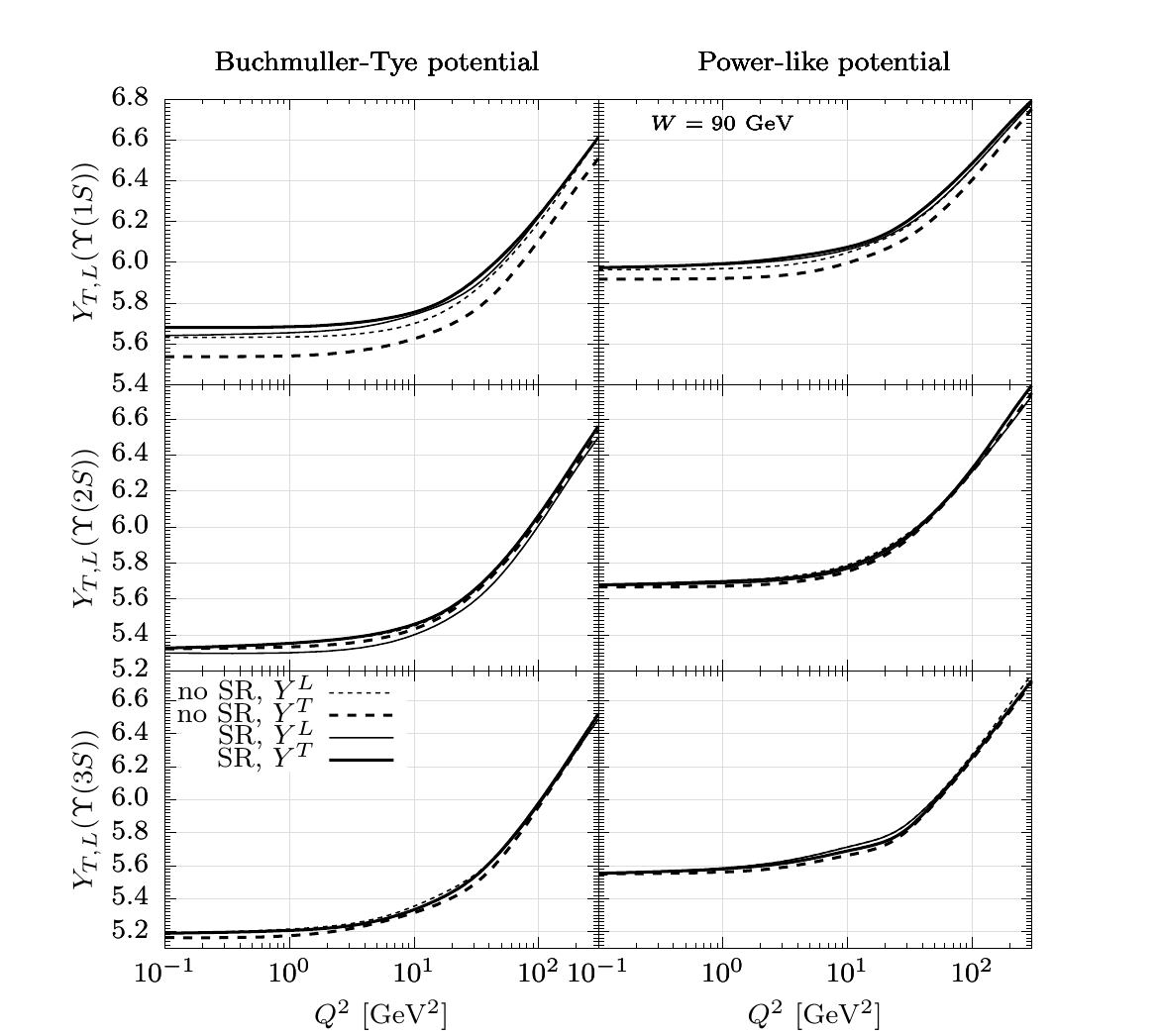}
    \caption{ \label{fig:upsilon-y}  
The $Q^2$ dependence of the scale parameters $Y_T$ and $Y_L$ defining the magnitude of the scanning radius, Eq.~(\ref{scan-rad}), for production of $T$ and $L$ polarized 
$\Y(1S)$ (upper boxes), $\Yp(2S)$ (middle boxes) and $\Ypp(3S)$ (lower boxes)
vector mesons. The solid and dashed curves represent the values of $Y_T$ and $Y_L$ with and without effects of the Melosh spin rotation, respectively. The thin and thick curves corresponds to production of $L$ and $T$ polarized vector mesons, respectively. The calculations have been performed with the vector meson wave function using the realistic Buchm\"uller-Tye \cite{bt-80} (left boxes) and power-low \cite{barik-80}
(right boxes) potential.
}
\end{center}
\end{figure}

However, as was already mentioned above, in the present paper following a justification in Ref.~\cite{yura-00} we take different LC quarkonium wave functions corresponding to different structure of vertices $\Psi\rightarrow Q\bar Q$ and $\gamma^*\rightarrow Q\bar Q$. Consequently, in comparison with the early results from Refs.~\cite{jan-94,jan-96,jan-07}, this leads to different expressions (\ref{AT}) and (\ref{AL}) for $T$ and $L$ production amplitudes and, consequently, to other relations between the factors $Y_T$ and $Y_L$ as is presented in Figs.~\ref{fig:psi-y} and \ref{fig:upsilon-y}.

At small $r_S\lsim r_0$ the scanning phenomenon \cite{bzk-93,bzk-94,jan-94,jan-96,jan-98,jan-01,jan-07}, Eq.~(\ref{scan-rad}), can be understood qualitatively by analyzing the forward production amplitudes (\ref{AT}) and (\ref{AL}), which can be evaluated as,
%
  \beqn
\mathrm{Im}\mathcal{A}_{T}
\propto 
r_S^2\,\sigma_{\bar qq}(r_S,s) \propto
\frac{Y_T^4}{(Q^2 + M_V^2)^2}\, ,
\label{rst}
  \eeqn
%
%
  \beqn
\mathrm{Im}\mathcal{A}_{T}
\propto \frac{\sqrt{Q^2}}{M_V}\,r_S^2\,
\sigma_{\bar qq}(r_S,s) & \propto &
\frac{\sqrt{Q^2}}{M_V}\,\frac{Y_L^4}{(Q^2 + M_V^2)^2}\, .
\label{rsl}
  \eeqn
%
%

Figs.~\ref{fig:psi-y} and \ref{fig:upsilon-y} show the $Q^2$-dependent scale parameters $Y_L$ and $Y_T$ for production of different heavy quarkonia using the BT and POW realistic potentials for determination of quarkonia wave functions. Evaluation
of $Y_L$ and $Y_T$ was performed from amplitudes (\ref{AT}) and (\ref{AL}) neglecting (dashed lines) and including (solid line) the Melosh spin rotation effects.

For electroproduction of bottomonia (see Fig.~\ref{fig:upsilon-y}) both scale parameters satisfy $Y_T\sim Y_L\sim 5.2\div 6.0$, depending on the shape of the $Q\bar Q$ interaction potential, and show a flat $Q^2$ behaviour as a consequence of the nonrelativistic approximation,
$z\sim 0.5$.
Besides, the onset of spin rotation effects does not lead practically to any changes of 
magnitudes for $Y_T$ and $Y_L$. One can also see that the scale parameters for production
of radially excited $\Yp(2S)$ and $\Ypp(3S)$ are slightly smaller than for $\Y(1S)$ as a manifestation of the node effect, i.e. due to a presence of nodes in the corresponding bottomonium wave functions for excited states. 

Analogical situation concerns also electroproduction of charmonia 
%
at $Q^2\lsim 10\div 20\,\GeV^2$,
%
when parameters $Y_{L}$ and $Y_{T}$ depicted in Fig.~\ref{fig:psi-y} exhibit a smooth rise with $Q^2$ and do not differ much from each other with the corresponding values similar to those for bottomonium production.
%
Here the integrands in Eqs.~(\ref{AT}) and (\ref{AL}) receive only a tiny contribution from asymmetric photon configurations, i.e. only the symmetric photon fluctuations with the meson momentum fraction $z\sim 0.5$ dominate. 
For this reason charmonium can be safely treated as a
nonrelativistic object at small and medium values of $Q^2$, when $r_S \gsim r_0$ (see Eq.~(\ref{pqcd})).
In comparison to production of $L$ polarized $\Jpsi(1S)$, the onset of spin rotation effects causes a slightly larger enhancement of the contribution from asymmetric configurations to the production of $T$ polarized $\Jpsi(1S)$ and consequently leads to a smooth inequality $Y_T(\Jpsi)\gsim Y_L(\Jpsi)$.
%
 
At larger $Q^2\gg M_{\Jpsi}^2$, the scale parameters $Y_{L,T}$ have a stronger $Q^2$ dependence reaching the values $\sim 7.0\div 7.2$ at $Q^2 = 100\,GeV^2$, which significantly differ from the nonrelativistic value $Y\lsim 6$. 
%
This is a direct consequence of the rising contribution from the large-size asymmetric $c\bar c$ photon fluctuations to the charmonium production.
%

Note that in comparison with bottomonium production, in electroproduction of charmonia
the difference in magnitudes between $Y_{L}$ and $Y_{T}$ is higher confirming the expected inequality, $Y_{T} > Y_{L}$, for $\Jpsi(1S)$ state. However, as 
a manifestation of the node effect, one can see a counter intuitive inequality, $Y_{L} > Y_{T}$, in production of radially excited $\psip(2S)$ state.

\begin{figure}[t]
\begin{center}
    \includegraphics[height=9.0cm]{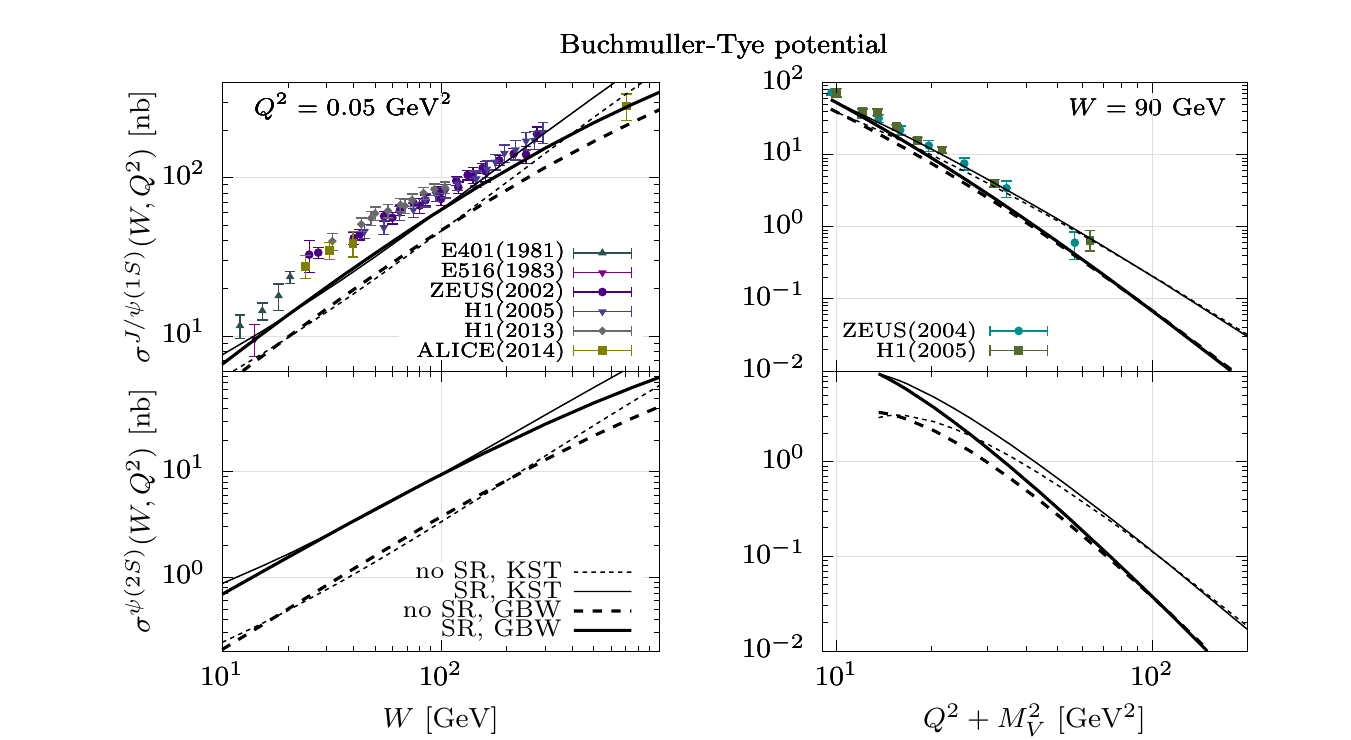}
    \caption{ \label{fig:bt-psi}  
Predictions for the elastic photoproduction cross section of the process $\gamma\,p\to \Jpsi(1S)\,p$ (left upper box) and $\gamma\,p\to \psip(2S)\,p$ (left lower box) with almost real photon ($Q^2 = 0.05\,\GeV^2$) as a function of c.m. energy $W$ of the photon-proton system. The corresponding data are taken from Refs.~\cite{e401,e516,zeus-02,h1-05,h1-13,alice-14}. Predictions for the elastic virtual photoproduction cross section of the process $\gamma^*\,p\to \Jpsi(1S)\,p$ (right upper box) and $\gamma^*\,p\to \psip(2S)\,p$ (right lower box) as a function of the photon virtuality $Q^2 + M_{\Jpsi}^2$ (right upper box) and $Q^2 + M_{\psip}^2$ (right lower box) at fixed c.m. energy $W=90\,\GeV$. The data of the process $\gamma^*\,p\to \Jpsi(1S)\,p$ are taken from Refs.~\cite{zeus-04,h1-05}. All model predictions have been performed with the wave functions of $\Jpsi(1S)$ and $\psip(2S)$ using the realistic Buchm\"uller-Tye potential \cite{bt-80}. The solid and dashed curves represent the model calculations with and without effects of the Melosh spin rotation, respectively. The thin and thick curves corresponds to calculations using the phenomenological KST \cite{kst-99} and GBW \cite{gbw} dipole cross section, respectively. 
}
\end{center}
\end{figure}

\begin{figure}[t]
\begin{center}
    \includegraphics[height=9.0cm]{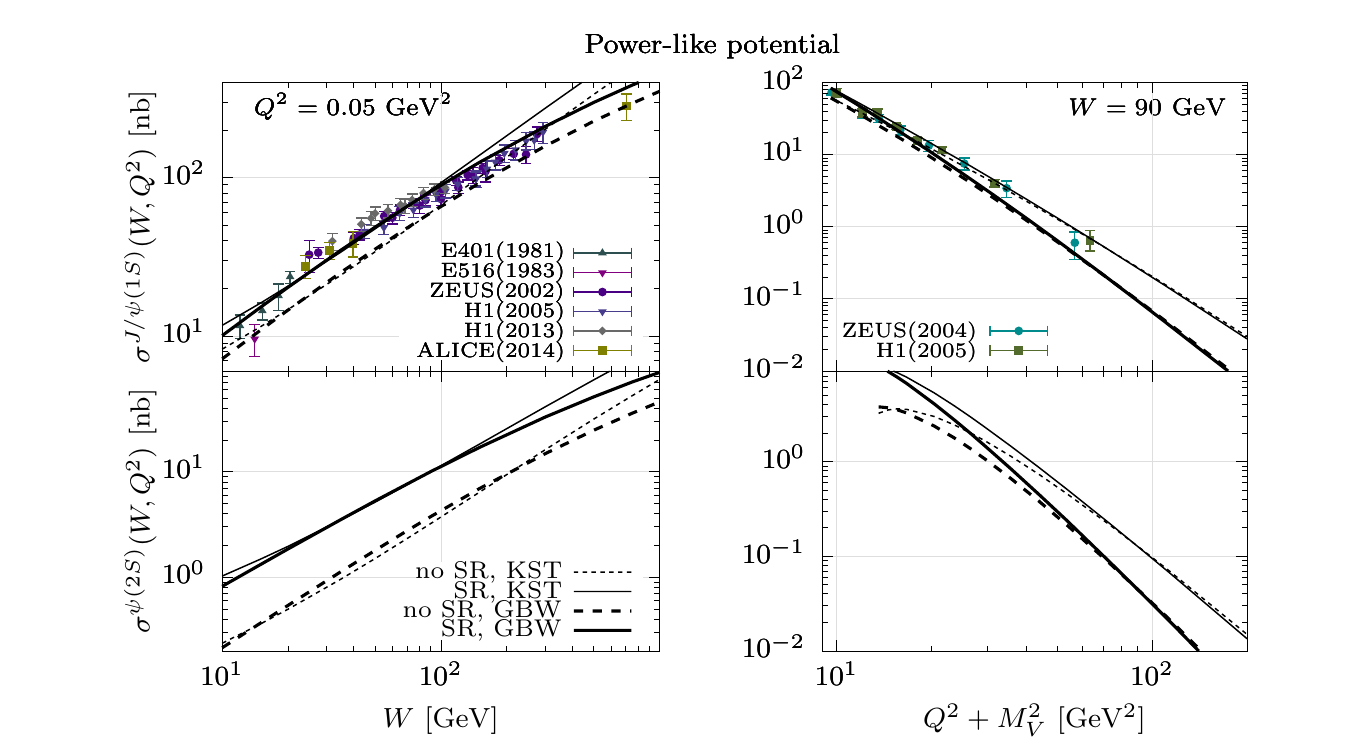}
    \caption{ \label{fig:pow-psi}  
The same as Fig.~\ref{fig:bt-psi} but with the wave functions of $\Jpsi(1S)$ and $\psip(2S)$ states obtained by using the realistic power-like potential \cite{barik-80}.
}
\end{center}
\end{figure}

As the next step, we analyse the onset of spin rotation effects in production of heavy
quarkonia. In Figs.~\ref{fig:bt-psi} and \ref{fig:pow-psi} we present the calculations
using Eq.~(\ref{final}) with amplitudes (\ref{AT}) and (\ref{AL}) for the integrated real (virtual) photoproduction cross section of the process $\gamma^*\,p\to \Jpsi(1S)\,p$
(upper boxes) and $\gamma^*\,p\to \psip(2S)\,p$ (lower boxes). The model predictions for
$\Jpsi(1S)$ state are compared with the data as functions of c.m. energy $W$ and $Q^2 + M_{\Jpsi}^2$. The latter represents the scaling variable as was discussed above. These calculations were performed with the KST and GBW parametrizations for the dipole cross section depicted by thin and thick lines, respectively. Besides, the Fig.~\ref{fig:bt-psi} and \ref{fig:pow-psi} corresponds to the charmonium wave functions calculated from the BT and POW potentials, respectively.

One can notice from Figs.~\ref{fig:bt-psi} and \ref{fig:pow-psi} an overall reasonable agreement of the model predictions with available data, whereas the calculations using the $\Jpsi(1S)$ wave function from the POW potential give a much better description. In the photoproduction limit ($Q^2\to 0$) such a good agreement with data is achieved mainly
due to effects of the Melosh spin rotation causing the significant $20-30\,\%$ enhancement of the photoproduction cross section. Otherwise, the model calculations underestimate the data by a factor of 1.5 especially at smaller c.m. energies $W\lsim 50\div 70\,\GeV$. At larger values of $Q^2$, the onset of spin rotation effects gradually diminishes with $Q^2$ as is demonstrated in right boxes of Figs.~\ref{fig:bt-psi} and
\ref{fig:pow-psi}.

Here, we would like to emphasize that inclusion of effects of the Melosh spin rotation is obviously neglected in recent studies of quarkonia produced in ultra-peripheral collisions at RHIC and LHC. Besides, a successful description of available data does not demand to include any additional new phenomena coming as another (hypothetical) reason for a required enhancement of the corresponding photoproduction cross sections.

In the case of photoproduction of the radially excited $\psip(2S)$ state, as a consequence of the node effect, both Figs.~\ref{fig:bt-psi} and \ref{fig:pow-psi} demonstrate that the onset of spin rotation effects is much stronger compared to $\Jpsi(1S)$ production increasing the photoproduction cross section by a factor of $2\div 3$.

There are other theoretical uncertainties which may affect a successful description of data, like sensitivity to heavy quark mass discussed in Ref.~\cite{yura-00}, the magnitude of the slope parameter in production of different heavy quarkonia and its dependence on the photon energy and virtuality, accuracy of the color dipole formalism at smaller c.m. energies $W\lsim 10\div 20\,\GeV$, inclusion of another realistic potentials for determination of the quarkonium wave functions and other more recent parameterizations for the dipole cross section, etc. Such a study is beyond the scope of the present paper, and will be presented elsewhere \cite{prepar}.

Universality in production of different heavy quarkonia states is controlled by the scanning radius Eq.~(\ref{scan-rad}). This leads to a conclusion that, for example, 
we expect the same onset of spin rotation effects as well as the same magnitudes of the cross sections for $\Jpsi(1S)$ electroproduction $\gamma^*\,p\to \Jpsi(1S)\,p$ at some scale $Q^2(\Jpsi)$, and for real photoproduction of $\Y(1S)$. The corresponding $Q^2(\Jpsi)$ can be estimated from Eq.~(\ref{scan-rad}) as follows,
%
\beqn
Q^2(\Jpsi) = 
M_{\Y}^2\,\frac{Y_{\Jpsi}^2}{Y_{\Y}^2} - M_{\Jpsi}^2\, ,
\label{jpsi-y}
\eeqn
%
where the scale factors $Y_{\Y}\sim 5.7$ and $Y_{\Jpsi}\sim 7.0$ for the BT potential can be extracted from Figs.~\ref{fig:psi-y} and \ref{fig:upsilon-y}. Thus, one arrives at the value $Q^2(\Jpsi)\sim 130\div 140\,\GeV^2$.
%
\begin{figure}[t]
\begin{center}
    \includegraphics[height=13.4cm]{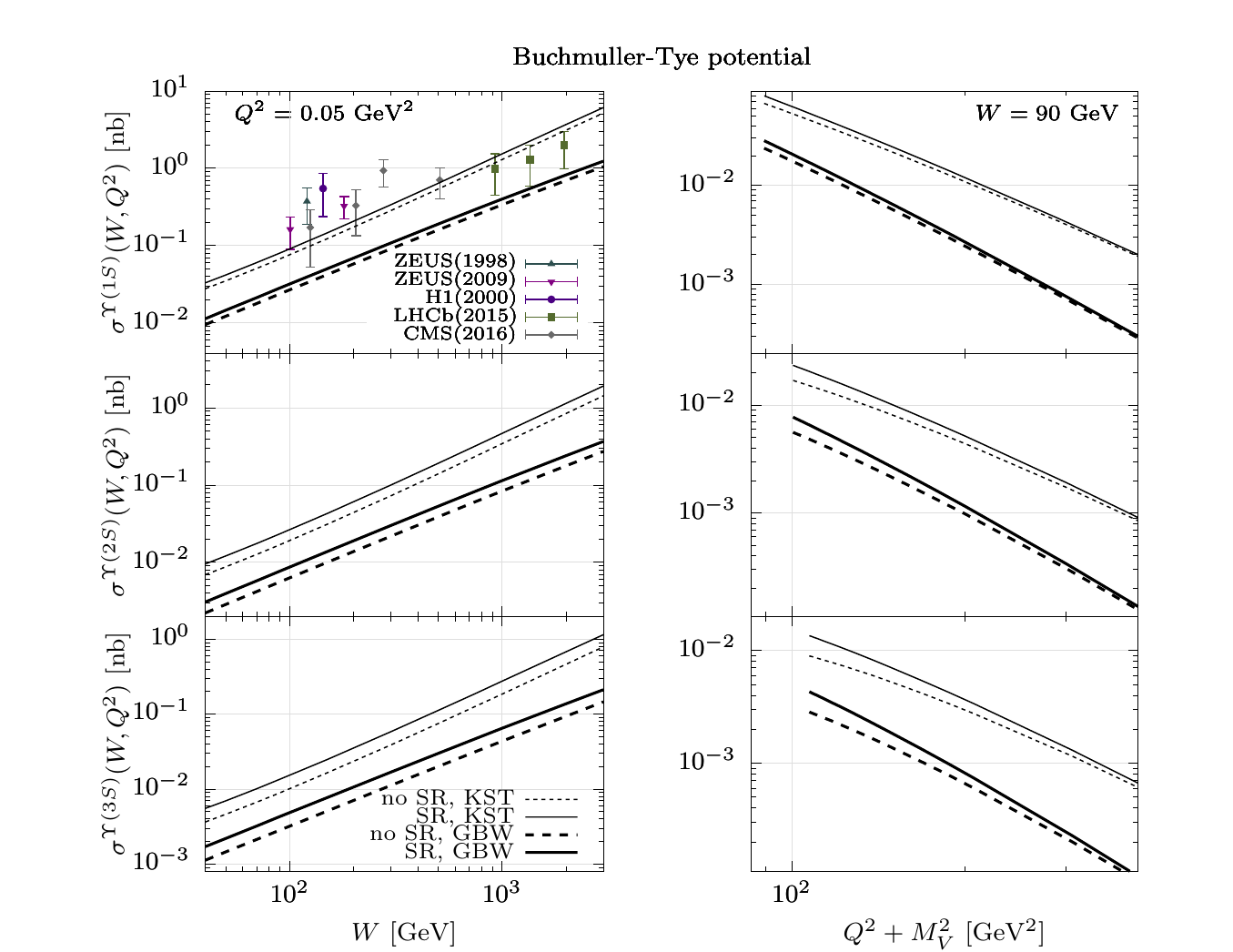}
    \caption{ \label{fig:bt-upsilon}  
The same as Fig.~\ref{fig:bt-psi} but for elastic virtual photoproduction of $\Y(1S)$ (upper boxes), $\Yp(2S)$ (middle boxes) and $\Ypp(3S)$ (lower boxes). The experimental data for the process $\gamma\,p\to \Y(1S)\,p$ are taken from Ref.~\cite{zeus-98,zeus-09,h1-00,lhcb-15,cms-16}. All model predictions have been obtained with the wave functions of $\Y(1S)$, $\Yp(2S)$ and $\Ypp(3S)$ using the realistic Buchm\"uller-Tye potential \cite{bt-80}.
}
\end{center}
\end{figure}

\begin{figure}[t]
\begin{center}
    \includegraphics[height=13.4cm]{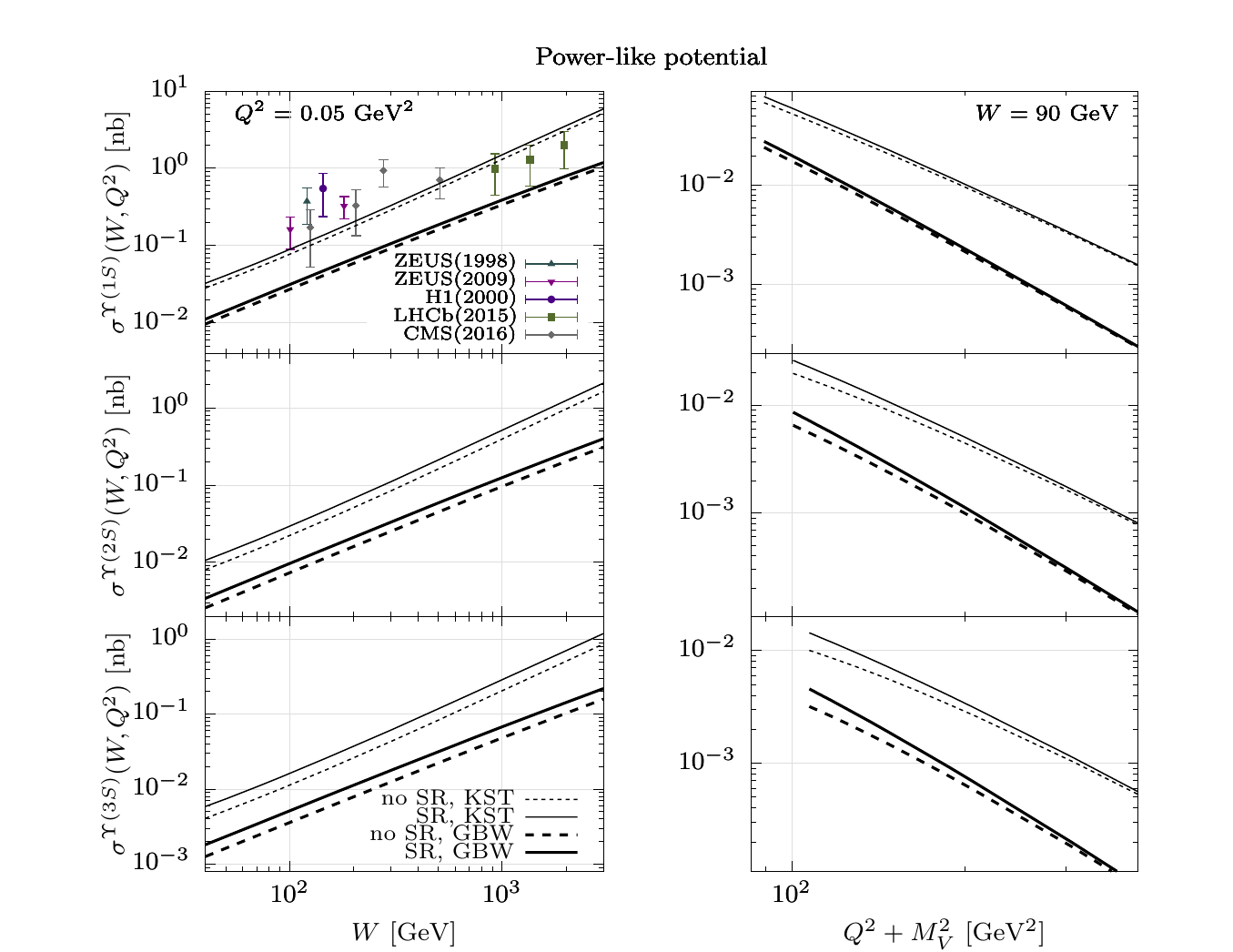}
    \caption{ \label{fig:pow-upsilon}  
The same as Fig.~\ref{fig:bt-upsilon} but with the wave functions of $\Y(1S)$, $\Yp(2S)$ and $\Ypp(3S)$ states obtained by using the realistic power-like potential \cite{barik-80}.
\vspace*{0.1cm}
}
\end{center}
\end{figure}

As we have mentioned above for electroproduction of charmonia, the spin rotation effects gradually vanish with $Q^2$. Thus one can expect, combining the scanning phenomenon (\ref{scan-rad}) with determination of the large scale $Q^2(\Jpsi)\sim 130\div 140\,\GeV^2$ from Eq.~(\ref{jpsi-y}), that the onset of spin rotation effects in $\Y(1S)$ photoproduction should be also very weak. Such an expectation is confirmed by Figs.~\ref{fig:bt-upsilon} and \ref{fig:pow-upsilon} for the BT and POW potentials,
where we present our model predictions for the c.m. energy $W$ and $Q^2$ dependence of the cross section for elastic photo- and electroproduction of bottomonia states $\Y(1S)$, $\Yp(2S)$ and $\Ypp(3S)$. One can also see that the onset of spin rotation effects
is stronger for radially excited states due to the presence of nodes in the corresponding wave functions.

For real photoproduction of $\Y(1S)$ our calculations are compared with available data. Upper left boxes of Figs.~\ref{fig:bt-upsilon} and \ref{fig:pow-upsilon} demonstrate a reasonable agreement, which is better for the KST than for the GBW dipole cross section.
As was already mentioned above, differences in predictions using the GBW and KST parametrizations can be treated as a measure of the underlined theoretical uncertainty.

\begin{figure}[t]
\begin{center}
    \includegraphics[height=5.6cm]{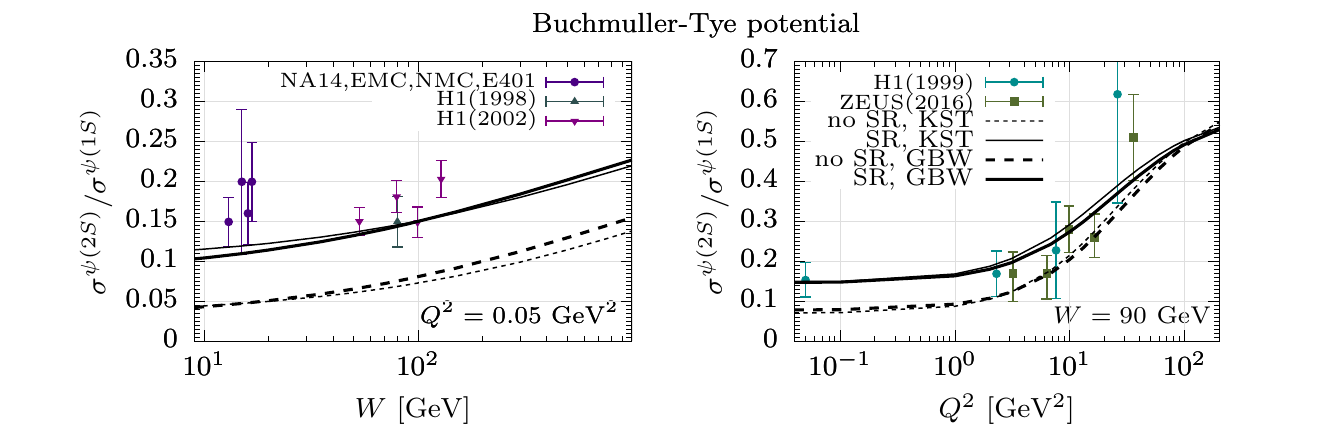} 
    \caption{ \label{fig:bt-rat-psi}  
Predictions for the ratio of $\psip(2S)$ to $\Jpsi(1S)$ photoproduction cross sections as a function of c.m. energy $W$ at fixed $Q^2 = 0.05\,\GeV^2$ (left box), and as a function of the photon virtuality $Q^2$ at fixed c.m. energy $W = 90\,\GeV$ (right box). All model predictions have been performed with the wave functions of $\Jpsi(1S)$ and $\psip(2S)$
using the realistic Buchm\"uller-Tye potential \cite{bt-80}. The solid and dashed curves represent model calculations with and without effects of the Melosh spin rotation, respectively. The thin and thick curves corresponds to calculations using the phenomenological KST \cite{kst-99} and GBW \cite{gbw} dipole cross sections, respectively. The experimental data points are taken from 
Refs.~\cite{na14-rat,emc-rat,nmc-rat,e401-rat,h1-98,h1-02,h1-99,zeus-16}. 
\vspace*{0.2cm}
} 
\end{center}
\end{figure} 

\begin{figure}[b]
\begin{center}
\vspace*{0.2cm}
    \includegraphics[height=5.6cm]{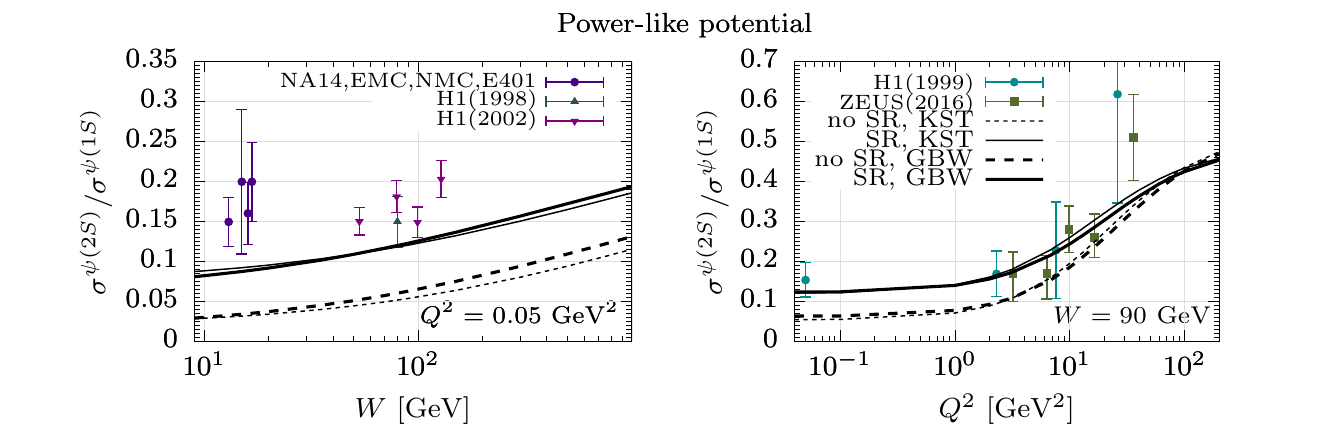} 
    \caption{ \label{fig:pow-rat-psi}  
The same as Fig.~\ref{fig:bt-rat-psi} but with the wave functions of $\Jpsi(1S)$ and $\psip(2S)$ states obtained by using the realistic power-like potential \cite{barik-80}.
}
\end{center}
\end{figure} 

As we have already analysed above, the spin rotation effects have a huge impact on the elastic photoproduction cross section of the $\gamma^*\,p\to \Jpsi(\psip(2S))\,p$  process. They add about $\sim 20\div 30\%$ to the $\Jpsi(1S)$ photoproduction cross section but cause a much more dramatic effect on $\psip(2S)$ state increasing the corresponding photoproduction cross section by a factor of $2\div 3$. Such a manifestation
of spin rotation effects can be tested investigating also the ratio $R$ of $\psip(2S)$ to $\Jpsi(1S)$ photo- and electroproduction cross sections as is depicted in Figs.~\ref{fig:bt-rat-psi} and \ref{fig:pow-rat-psi} for the BT and POW potentials as functions of c.m. energy $W$ and $Q^2$. In comparison with the standard photoproduction
cross section the study of $R$ allows to minimize the theoretical uncertainties connected mainly with our choice of parameterization for the dipole cross section, with determination of the corresponding slope parameters for the considered processes
$\gamma^*\,p\to \Jpsi(\psip(2S))\,p$ and quarkonium wave functions etc.

One can see a reasonable agreement of our calculations with available data. For a more reliable test of the model one needs more precise data. The rise of the ratio $R(W)$
with energy is in variance with the natural expectation that, in comparison to $\Jpsi(1S)$, the larger size of $\psip(2S)$ should lead to a weaker dependence on energy since smaller $Q\bar Q$ dipole sizes have a steeper rise with energy. However, the predicted rise of $R(W)$ is another manifestation of the node effect. The small-size
part of the $\psip(2S)$ wave function below the node position has a steeper energy dependence leading to a strong reduction of a cancellation in the production amplitude. As a result, the energy dependence of $\psip(2S)$ production is steeper compared to that of $\Jpsi(1S)$. Similarly, the node effect causes a rise of the ratio $R(Q^2)$ with $Q^2$ as is shown in right boxes of Figs.~\ref{fig:bt-rat-psi} and \ref{fig:pow-rat-psi} where our predictions for the BT and POW potentials are compared with available data.

\begin{figure}[t]
\begin{center}
    \includegraphics[height=8.5cm]{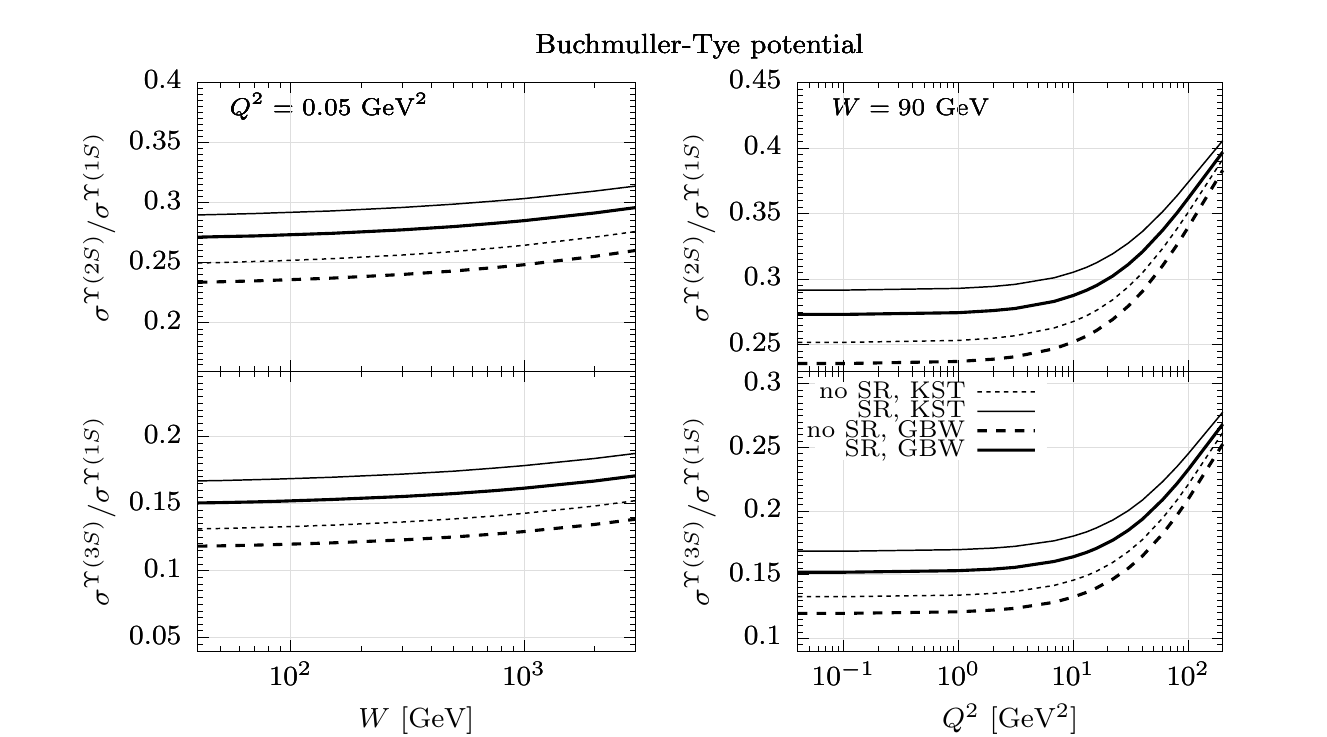} 
    \caption{ \label{fig:bt-rat-ups}  
The same as Fig.~\ref{fig:bt-rat-psi} but for the ratio of $\Yp(2S)$ to $\Y(1S)$ (upper boxes) and $\Ypp(3S)$ to $\Y(1S)$ (lower boxes) photoproduction cross sections.
All model predictions have been performed with wave functions of $\Y(1S)$, $\Yp(2S)$ and $\Ypp(3S)$ states obtained by using the realistic Buchm\"uller-Tye potential \cite{bt-80}.
} 
\end{center}
\end{figure} 
\begin{figure}[b]
\begin{center}
    \includegraphics[height=8.5cm]{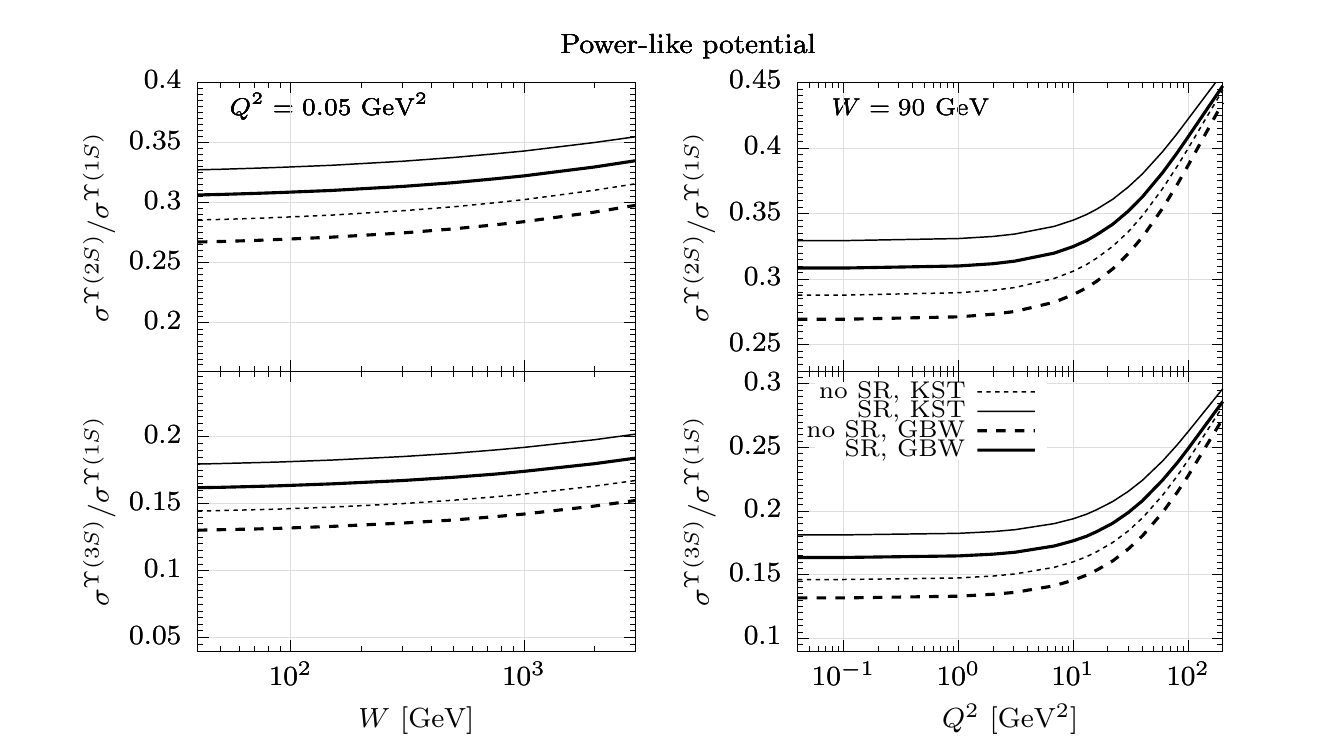} 
    \caption{ \label{fig:pow-rat-ups}  
The same as Fig.~\ref{fig:bt-rat-ups} but with the wave functions of $\Y(1S)$, $\Yp(2S)$ and $\Ypp(3S)$ states obtained by using the realistic power-like potential \cite{barik-80}.
} 
\end{center}
\end{figure} 

In the last two Figs.~\ref{fig:bt-rat-ups} and \ref{fig:pow-rat-ups} we present our predictions for the ratios of $\Yp(2S)$ to $\Y(1S)$ and $\Ypp(3S)$ to $\Y(1S)$ photo-
and electroproduction cross sections as function of c.m. energy $W$ and photon virtuality $Q^2$ for the BT and POW potentials. As was discussed above, the smaller $Q\bar Q$ dipole sizes at larger scale $Q^2 + M_V^2$ diminish the spin rotation effects. Consequently,
in comparison to $\Y(1S)$, the larger sizes of radially excited bottomonia cause
a stronger onset of these effects, whereas the higher size of $\Ypp(3S)$ compared to
$\Yp(2S)$ leads even to an additional enhancement of the corresponding photoproduction cross section.

Figs.~\ref{fig:bt-rat-ups} and \ref{fig:pow-rat-ups} clearly demonstrate the importance of the Melosh spin rotation effects increasing significantly the ratios 
$\sigma^{\gamma^*\,p\to\Yp(2S)\,p}/\sigma^{\gamma^*\,p\to \Y(1S)\,p}$
and
$\sigma^{\gamma^*\,p\to \Ypp(3S)\,p}/\sigma^{\gamma^*\,p\to \Y(1S)\,p}$.
As was mentioned above, the differences in predictions employing the GBW and KST parametrizations can be treated as a measure of the theoretical uncertainty in the current analysis.

\section{Conclusions}
\label{Sec:conclusions}

In this paper, we present for the first time the detailed exploratory study of the Melosh spin rotation effects in elastic electroproduction $\gamma^*\,p\to V\,p$ of $S$-wave heavy quarkonia, where $V = \Jpsi(1S), \psip(2S), \Y(1S), \Yp(2S)$ and $\Ypp(3S)$.

The final factorized light-cone expressions for transversely (\ref{AT}) and longitudinally (\ref{AL}) polarized production amplitudes are based on our knowledge of the following ingredients: 
(i) the universal dipole cross section $\sigma_{q\bar q}(r,s)$ which depends on the transverse separation $r$ between $Q$ and $\bar Q$ and the c.m. energy squared, and describes the interaction of the $Q\bar Q$ dipole with a proton target;
(ii) the perturbative light-cone wave function for the $Q\bar Q$ component of the
photon $\Psi_{\gamma^*}(r,z;Q^2)$ which depends on the photon virtuality $Q^2$ and the relative fraction $z$ of the photon momentum carried by the quark; and
(iii) the light-cone wave function $\Psi_V(r,z)$ for the $S$-wave heavy quarkonium.

In our calculations of the photo- and electroproduction cross sections of heavy quarkonia, including also the effects of the spin rotation, we adopted as a reference 
the two distinct parametrizations for the dipole cross section $\sigma_{q\bar q}(r,s)$
in the saturated form at large $r$ and satisfying the small-$r$ behavior $\sigma_{q\bar q}(r)\propto r^2$ (color transparency), with the fitted parameters to obtain the best description of data for $\sigma^{\pi p}_{tot}(s)$ and the structure functions $F_2(x,Q^2)$. The dipole cross section at smaller $Q\bar Q$ transverse separations has a steeper rise with energy.

There are no large uncertainties in description of the LC photon wave function.
The same is not true for the light-cone wave functions of quarkonia where we are forced to use the standard prescription realizing the transition from a nonrelativistic wave
function in the $Q\bar Q$ rest frame to its LC form in the light-front description. The former has been obtained by solving the Schroedinger equation for two different heavy-quark interaction potentials, which provide the best description of the available data on $\Jpsi(1S)$ and $\Y(1S)$ photo- and electroproduction, as well as on the ratio of $\psip(2S)$ to $\Jpsi(1S)$ photoproduction cross sections.

Universality in production of different heavy quarkonia states is controlled by the scanning radius Eq.~(\ref{scan-rad}), and leads to very similar magnitudes of electroproduction cross sections at fixed values of the scaling variable $Q^2+M_V^2$.
However, for production of radially excited heavy quarkonia this scanning phenomenon should be treated at sufficiently large $Q^2$ when the corresponding scanning radius $r_S\ll r_n$, with $r_n$ being the position of the first node in the quarkonium
wave function.

In our model predictions we have included the Melosh spin rotation which is often neglected in present calculations of the corresponding cross sections for elastic processes $\gamma^*\,p\rightarrow V\,p$, where $V = \Jpsi, \psip, \Y, \Yp, \Ypp, ..$.
Our observations are the following: \\
(i) 
Universal properties in production of different vector mesons controlled by the scanning radius Eq.~(\ref{scan-rad}) lead to very similar magnitudes of the Melosh spin rotation effects in production of charmonia and bottomonia at the same fixed values of $Q^2+M_V^2$;\\
(ii) 
Spin rotation effects contribute to obtain a reasonable agreement with available data
without any adjusted parameter;\\ 
(iii)
Spin rotation effects lead to a rise of the $\psip(2S)$ photoproduction cross section by a factor of $2\div 3$ causing a substantial enhancement of the $\psip(2S)$ to $\Jpsi(1S)$ ratio of the photoproduction cross sections to the values close to experimental data;\\
(iv)
Spin rotation effects are gradually diminished with $Q^2$, and we also find a weak onset of these effects in photoproduction of the $\Y(1S)$ state as a manifestation of the scanning phenomenon;\\
(v)
Similarly to production of radially excited charmonia, we predict a stronger onset of the spin rotation effects in production of $\Yp(2S)$ and $\Ypp(3S)$ states as a direct manifestation of the nodal structure of their wave functions. Notably, the spin rotation effects in production of $\Ypp(3S)$ are even stronger compared to $\Yp(2S)$ production.

In conclusion, a large importance of the Melosh spin transformation, especially in photoproduction of heavy quarkonia, can contribute to a better understanding of dynamics in production of charmonia and bottomonia in ultra-peripheral $pA$ and heavy-ion collisions at RHIC and LHC, and thus should also be taken into account and tested by future measurements at electron-ion colliders.

\section*{Acknowledgements}

J.N.~is partially supported by grants LTC17038 and LTT18002
of the Ministry of Education, Youth and Sports of the Czech Republic
and
by projects of the European Regional Development Fund
CZ02.1.01/0.0/0.0/16\_013/0001569 and CZ02.1.01/0.0/0.0/16\_019/0000778.
R.P.~is supported in part 
by the Swedish Research Council grants, contract numbers 621-2013-4287 and 
2016-05996, by CONICYT grant MEC80170112, by the Ministry of Education, Youth and 
Sports of the Czech Republic, project LT17018, as well as by the European Research 
Council (ERC) under the European Union's Horizon 2020 research and innovation 
programme (grant agreement No 668679). The work of M.K. was supported in part by 
the Conicyt Fondecyt grant Postdoctorado N.3180085 (Chile) and by the grant LTC17038 
of the Ministry of Education, Youth and Sports of the Czech Republic. The work has 
been performed in the framework of COST Action CA15213 ``Theory of hot matter 
and relativistic heavy-ion collisions'' (THOR). 

\bibliographystyle{unsrt}

\end{document}